\documentclass[aps,pre,preprint,showpacs,groupedaddress]{revtex4-1}

\usepackage{graphicx}
\usepackage{amsmath}
\usepackage{amsfonts}
\usepackage{comment}
\usepackage{color}

\begin{document}

%\preprint{}

\title{Derivation of the Biot-Savart equation from the Nonlinear Schr\"odinger equation}



\author{Miguel D. Bustamante}
\email[Corresponding author: ]{Miguel.Bustamante@ucd.ie}
\affiliation{Complex and Adaptive Systems Laboratory, School of Mathematics and Statistics, University College Dublin, Belfield, Dublin 4, Ireland}

\author{Sergey Nazarenko}
\email{S.V.Nazarenko@warwick.ac.uk}
\affiliation{Mathematics Institute, The University of Warwick, Coventry, CV4 7AL, United Kingdom}

%\thanks{}
%\affiliation{}


\date{\today}

\begin{abstract}
We present a systematic derivation of the Biot-Savart equation from the Nonlinear Schr\"odinger  equation, in the limit when the curvature radius of vortex lines and the inter-vortex distance are much greater than the vortex healing length, or core radius.
We derive the Biot-Savart equations in Hamiltonian form with Hamiltonian expressed in terms of vortex lines, 
$$ H= \frac{\kappa^2}{8 \pi}\int_{|{\bf  s}-{\bf  s}'|>\xi_*} \frac{d{\bf  s} \cdot d {\bf  s}'}{|{\bf  s}-{\bf  s}'|} \,,$$
with  cut-off length   $\xi_* \approx 0.3416293 /\sqrt{\rho_0},$ where $\rho_0$ is the background condensate density far from the vortex lines and $\kappa$ is the quantum of circulation.

\end{abstract}


\maketitle


\section{Introduction}

Nonlinear Schr\"odinger (NLS) equation 
\eqref{eq:NLS} and Biot-Savart equation 
\eqref{Biot-Savart equation} are the two most popular
models for describing superfluid dynamics and turbulence
at very low temperature \cite{Barenghi01-proj, koplik,Tsubota2000,Tsubota2002,norePRL97, norePF97,Nemirovskii2015,Nemirovskii2004}.
This includes flows in superfluid $^4$He and $^3$He, as well
as atomic Bose-Einstein condensates of alkali gases.
It is generally believed that the Biot-Savart equation describes
a subset of slow (subsonic) motions of the more general 
NLS   equation.
This view is based on the formal ideal-fluid analogy in the NLS model
arising from the Madelung transformation, as will be explained below in
Section \ref{madelung}.
However, to date there has been no rigorous justification for such
a correspondence between the Biot-Savart and the
NLS   equations.
The problem is that the zero vortex line radius limit is ill defined
for the ideal fluids leading to a divergence of the Biot-Savart integral.
To tackle this problem, it is customary to introduce a phenomenological
cut-off regularisation of the Biot-Savart integral at a scale that corresponds roughly
 to the vortex line radius.
All efforts to justify this approach rigorously for the ideal fluid dynamics
have failed due to a great freedom in possible realisations of the vortex profiles and their temporal variability due to vortex stretching.

Fortunately, the fluid dynamics equations arising from the NLS equation
contain an extra term -- the so called quantum pressure (see Section \ref{madelung}). It is the quantum pressure term that makes the quantum vortex core ``rigid", i.e. having a fixed universal profile. This fact makes it possible to rigorously derive the Biot-Savart equation from the NLS equation. The present paper is devoted to such a derivation.

\section{Nonlinear Schr\"odinger equation}

Consider a model described by the 3D defocusing
Nonlinear Schr\"odinger  (NLS)  equation \cite{gross, pitaevskii1961vortex},
{\begin{equation}
\nonumber
i \hbar \, \partial_T \Psi + \frac{\hbar^2}{2 M}\nabla_{\textbf{X}}^2 \Psi + E \Psi - V_0 |\Psi|^2 \Psi 
=0\,,
\end{equation}
where $\Psi$ is a complex scalar function of a 3D physical space coordinate $\textbf{X}$ and time $T$, $\hbar$ is the reduced Planck's constant, $M$ and $E$ are the mass and single-particle energy of the bosons, and $V_0$ is the strength of the interaction potential between them.  Define dimensionless variables as follows: $\textbf{x} = \frac{\sqrt{2ME}}{\hbar} \textbf{X}, \,\, t =\frac{E}{\hbar}\, T$ and $\psi = \sqrt{\frac{V_0}{E}}\, \Psi$. The resulting equation is:}
\begin{equation}
\label{eq:NLS}
	i \, \partial_t \psi + \nabla^2 \psi + \psi - |\psi|^2 \psi 
=0 \,.
\end{equation}

\subsection{Hamiltonian  formulation}

The  NLS equation (\ref{eq:NLS}) can be written in Hamiltonian form,
\begin{equation}
\label{eq:NLS-h}
	i \, \partial_t \psi = \frac{\delta H}{\delta \psi^*}
\ ,
\end{equation}
with Hamiltonian
{\begin{equation}
H =\int \left[ |\nabla \psi(\mathbf{x}, t)|^2 + \frac{1}{2}\left(|\psi(\mathbf{x}, t)|^2-1\right)^2 \right] d\mathbf{x} \, ,
\label{eq:cons-E}
\end{equation}}
which represents the conserved energy of the system.

\subsection{Madelung transformation and fluid framework}
 \label{madelung}

The Madelung transformation 
\cite{Madelung,Spiegel,norePF97}
maps the complex scalar field $ \psi(\mathbf{x}, t) $ to two real scalar fields $ \rho(\mathbf{x}, t) $ and $ \phi(\mathbf{x}, t) $ following the relation $ \psi = \sqrt\rho \, e^{i\phi} $.
By plugging this substitution into the defocusing NLS equation (\ref{eq:NLS}) and separating the real and imaginary parts, one obtains the set of equations
{\begin{equation*}
\begin{split}
& \frac{\partial \rho}{\partial t} + \nabla \cdot (\rho \, 2 \nabla\phi) =0, \\
& \frac{\partial \phi}{\partial t} +  (\nabla \phi )^2 + \rho - 1 -\frac{\nabla^2 \sqrt{\rho}}{\sqrt{\rho}} =0.
\end{split}
%\label{eq:madelung-1}
\end{equation*}}
It is then straightforward to observe that by setting the vector field $ \mathbf{u}(\mathbf{x}, t) = 2 \nabla \phi(\mathbf{x}, t) $ the first equation results in a continuity equation for the density field $ \rho $ and the second equation a conservation of the  momentum associated with the velocity field $ \mathbf{u} $:
\begin{equation*}
\begin{split}
& \frac{\partial \rho}{\partial t} + \nabla \cdot (\rho \, \mathbf{u}) =0, \\
& \frac{\partial \mathbf{u}}{\partial t} +  (\mathbf{u} \cdot \nabla) \, \mathbf{u} = - \frac{\nabla \rho^2}{\rho} + \nabla \left(2\frac{\nabla^2 \sqrt{\rho}}{\sqrt{\rho}}\right).
\end{split}
%\label{eq:madelung-2}
\end{equation*}
Thus, we have obtained equations of an inviscid polytropic gas with adiabatic index $\gamma=2$ (pressure $p=\rho^2$), but with an  extra term, called quantum pressure, appearing at the last term in the second equation. The quantum pressure term is negligible if the characteristic scale of motion is much greater than the healing length $\xi$.

Let us re-express  Hamiltonian \eqref{eq:cons-E} in terms of the fluid variables:
{\begin{equation}
H =\frac{1}{2} \int \left[ 
  \frac{1}{2} \rho {u}^2
+  (\rho-1)^2  + 2|\nabla \sqrt{\rho}|^2 \right]  \, d\mathbf{x}  \, .
\label{eq:cons-Ef}
\end{equation}}
With the exception of the last term,  we see a formal coincidence  with the standard expressions for the  total energy (up to  factor $1/2$)  of a compressible fluid.

%%%%%%%%%%%%%%%%%%%%%%%%%%%%%%

\section{Quantised vortices}

\label{sec:Quantised vortices}

Even if the velocity field $ \mathbf{u} $ is irrotational, vortices may appear in the system.
These are lines of singular vorticity around which the real phase field changes by a multiple of $ 2\pi $ assuring that the complex wave field $ \psi $ stays always single-valued.
One can measure the circulation along a closed curve $ C $ around one of these vortex lines
\begin{equation*}
\Gamma = \oint_C \mathbf{u} \cdot d\mathbf{l} = 2 \oint_C \nabla \phi \cdot d\mathbf{l} = 2  \, \Delta \phi =  4\pi  n \, , \quad \mbox{with} \quad n \in \mathbb{Z} \, .
%\label{eq:circulation}
\end{equation*}
The circulation is zero for all contours embracing the regions  where the phase is a well-defined  differentiable function. However,  at points where $\psi =0$, the phase is undefined and the circulation along contours embracing such points is  not zero.
The circulation may take only discrete values and the singularities in the vorticity field are called quantum vortices. Moreover, vortices with 
$n \ge 2$ are unstable: a general smooth change in field $\psi$ leads to splitting of such ``multi-charge'' vortices into a set of elementary vortices with $n=1$.
Because of the quantised circulation, together with the fact that the fluid has no viscosity, the NLS  model has been used to qualitatively describe superfluids.

Let us first consider a  single straight vortex solution found by Pitaevskii \cite{pitaevskii1961vortex}.
For simplicity, let us consider a situation where the density far from vortices asymptotes to $\rho_0 =1$. The more general case when the density tends to a different constant $\rho_0 >0$ can then be obtained by a simple rescaling $\psi ({\bf x}, t) \to \sqrt{\rho_0} \, \psi ( \sqrt{\rho_0} \, {\bf x}, t)$.
Let us impose that a phase shift of $ +2\pi $ exists around the origin for the phase field $ \phi $ and look for a stationary solution.
Going into polar coordinates
\begin{equation*}
\left\{
\begin{split}
& x = r \cos\theta \\
& y = r \sin\theta
\end{split}
\right.
\end{equation*}
we consider a solution with   {$ \phi(r, \theta) = \theta$.}
We will also impose an axial symmetry on the density field, so that
the solution has the form
{\begin{equation}
\label{pv}
\psi_v(r, \theta) = R(r) \, e^{i \theta}.
\end{equation}}
Plugging this  into the  NLS equation (\ref{eq:NLS}) we get the ordinary differential equation
\begin{equation}
\frac{d^2 R}{dr^2} + \frac{1}{r} \frac{d R}{d r} - \frac{1}{r^2} R + \left(1-R^2 \right) R = 0 \, ,
\label{eq:NLS-vortex}
\end{equation}
supplemented with the boundary conditions $R(0) = 0, \quad R(\infty) = 1.$ One may obtain analytically the behaviour of the function $ R(r) $ in the limits of very small and very large radius, or look for a numerical solution using a shooting method. Another method is to use a Pad{\'e} approximation \cite{berloff:2004}. For the present paper, it turns out, the computation of the cut-off length $\xi_*$ (see Section \ref{sec:Rigorous}) requires  knowledge of a very accurate solution for $R(r)$ over a wide range of values of $r$. This rules out Pad\'e approximations (they have significant errors of up to $3.5\%$) so we are forced to work with high-accuracy numerical solutions and asymptotic solutions at large $r$.  Figure \ref{fig:1} shows a log-linear plot of $R(r)$ obtained by our method as explained in Appendix \ref{sec:accu_R}. The vortex core has size of the order of the healing length tending rapidly to a constant value far from the origin. 
The field $ \psi $ is smooth everywhere  and tends to zero at the origin which can be thought of as a localised topological phase defect.

\begin{figure}[h]
\includegraphics[width=9cm]{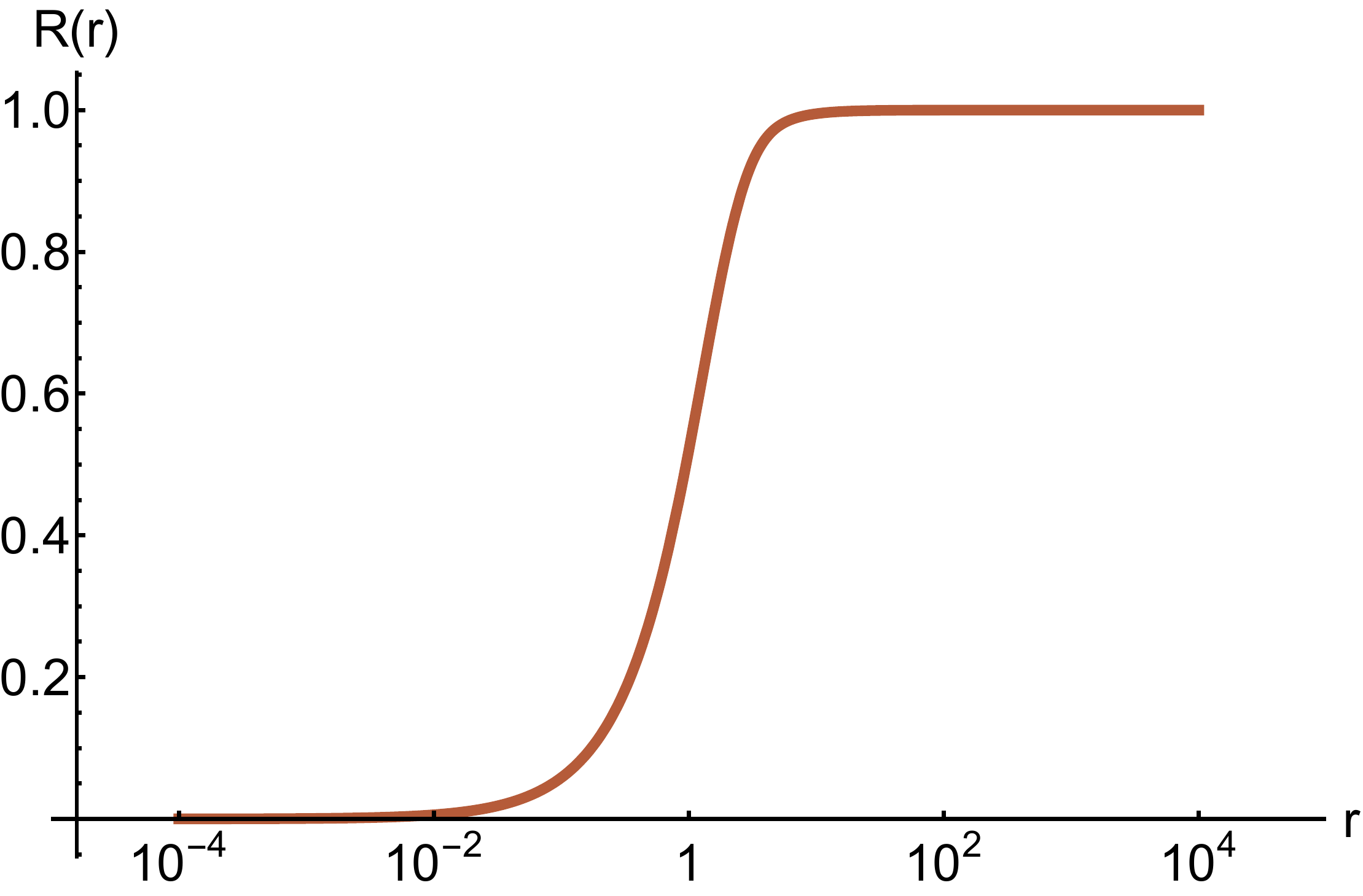}
\caption{
\label{fig:1}
(Color online) Log-linear plot of the vortex profile $R(r)$, obtained as a piecewise combination between a highly-accurate numerical solution and an asymptotic solution. See Appendix \ref{sec:accu_R} and Supplemental Material \cite{suppl_mat} for details.}
\end{figure}

\section{Vortex tangle and Biot-Savart model}

General vortex structure in strong 3D NLS turbulence may be very complex and irregular: it is usually referred as {\em vortex tangle}.
In a wider context, vortex tangle represents a typical realisation of superfluid turbulence at zero temperature, i.e. in liquid Helium 4 or 3.

%\subsection{Biot-Savart model: Phenomenology and heuristic derivations} 

When the distance between the vortex lines and their curvature radii are much greater than the healing length $\xi$, 
the first term in the integral of energy \eqref{eq:cons-Ef}, the so-called kinetic energy, is dominant over the second and the third contributions, the internal and the quantum-pressure energies respectively.
This is because (provided there is no ambient sound) the density field deviations from the background density $\rho_0$ (in our case $\rho_0=1$) are bound within small vortex cores of radius $\sim\xi$, whereas the velocity field $\mathbf{u}$ produced by the vortices is de-localised. Thus, the leading order in the Hamiltonian is
\begin{equation}
H =\frac{1}{4} \int 
  \rho {u}^2
 \, d\mathbf{x}  \, ,
\label{eq:cons-Ebs}
\end{equation}
which is a Hamiltonian of an incompressible fluid (up to factor $1/2$), but with density depletions in the vortex cores which are rigidly fixed to have transverse density distributions of the isolated 2D vortex considered in Section
\ref{sec:Quantised vortices}. (Note for our purposes 
we will also need to find the main sub-leading order in $H$). 

This suggests that
the vortex tangle can be modelled by the Biot-Savart equation, which has been used the context of the vortex dynamics in incompressible fluids starting with the work of Da Rios \cite{DaRios}:
\begin{equation}\label{Biot-Savart equation}
{\bf  s}_t \equiv 
 \frac{\partial   {\bf  s}}{\partial   t} = \frac{\kappa}{4\pi}\int \frac{({\bf  s}'-{\bf  s}) \times d{\bf  s}'}{|{\bf  s}-{\bf  s}'|^3}\ ,
\end{equation}
where $ {\bf  s} \equiv  {\bf  s} (\zeta, t)$, where $\zeta \in \mathbb R $ is a Lagrangian label parametrising the positions of the vortex line elements.  Here, $\kappa$ is the quantum of circulation; in the non-dimensional NLS model considered in this paper we have $\kappa = 4 \pi$.
The integral is taken along all of the vortex lines, including the one to which the considered vortex element belongs.
 To prevent logarithmic  singularity at ${\bf  s}' \to {\bf  s}$,  the integration has a cut-off   at the ``vortex radius'' scale,
$\xi_* < |{\bf  s}-{\bf  s}'|$, whose value, based on the common sense physical grounds, must be close to the healing length $\xi$. Such a cut-off has been previously introduced
phenomenologically, but later in the present paper we will provide a
rigorous justification for this model and will find the value
of  $\xi_*$ numerically.

Equation \eqref{Biot-Savart equation} can be obtained from the least-action principle
$$ \delta S = \int \delta {\cal L} \, dt = 0 $$
with Lagrangian 
\cite{Rasetti,Kuznetsov2000,holm2004}:
\begin{equation*}%\label{L Biot-Savart equation}
 {\cal L} = \frac \kappa 6 \int {\bf s}_t \cdot ({\bf s} \times  d {\bf s}) - {\cal H}
\end{equation*}
and Hamiltonian
\begin{equation}\label{ham Biot-Savart equation}
 {\cal H} = \frac{\kappa^2}{16\pi}\int
%_{|{\bf  s}'-{\bf  s}''| > \xi} 
\frac{  d{\bf  s}' \cdot d{\bf  s}''}{|{\bf  s}'-{\bf  s}''|}\ .
\end{equation}

Variation of action
gives:
\begin{equation*}%\label{S  var Biot-Savart equation}
\delta S = -\int dt  \, \delta {\bf s} \cdot \left[
\frac \kappa 2 \, {\bf s}_t \times  {\bf s}_\zeta + \frac{\delta {\cal H}}{\delta {\bf s}} 
 \right] d\zeta =0.
\end{equation*}
Considering the fact that the parametrisation $\zeta$ is arbitrary, the expression in the square brackets must be zero for any
vector $ {\bf s}_\zeta$ tangential to $ {\bf s}$.
The respective Hamiltonian equation is:
\begin{equation}\label{Ham BS equation}
\frac \kappa 2  \, {\bf s}_t \times  {\bf s}_\zeta = - \frac{\delta {\cal H}}{\delta {\bf s}}.
\end{equation}
Taking into account that
\begin{equation*}%\label{Ham der BS }
\frac{\delta {\cal H}}{\delta {\bf s}} =
\frac{\kappa^2}{8\pi} \int  \frac{   
({\bf  s}_\zeta \cdot ({\bf  s}-{\bf  s}'))\,  d{ \bf  s}' - ({\bf  s}_\zeta \cdot d{\bf  s}') ({  \bf s}-{\bf  s}')   }{|{\bf  s}-{\bf  s}'|^3}
=  \frac{\kappa^2}{8\pi} \int  \frac{   {\bf  s}_\zeta \times
[({\bf  s}'-{\bf  s}) \times  d{ \bf  s}']}{|{\bf  s}-{\bf  s}'|^3}
\end{equation*}
and undoing the cross product, we get the Biot-Savart equation
\eqref{Biot-Savart equation}.

\section{Derivation of the Biot-Savart model from the NLS equation} 

\label{sec:Rigorous}

In spite of great popularity of the Biot-Savart model, it has not yet been rigorously obtained or justified in the form formulated here, i.e. with an integration cut-off.
In the ideal fluids context, the difficulty is related to the vortex stretching effect, which causes changes in the vortex core radius.
The situation may be easier in the NLS model, since the vortex core shape is fixed. However, previous attempts to derive the vortex line evolution in the NLS model have led to significantly more complex equations than equation \eqref{Biot-Savart equation}  (see e.g.
\cite{pismen1999vortices}).

The present paper is devoted to such a derivation of the Biot-Savart model within the NLS model. For this one has to (A) rewrite the NLS
equation in the vortex filament form \eqref{Ham BS equation}, and (B)
express the Hamiltonian $H$ in terms of the vortex line configuration
${\bf s}(\zeta,t)$. Part (A) was done in \cite{Nemirovskii2004} in more general 
settings (forced Ginzburg-Landau equation), and we reproduce it 
for our NLS case in the Appendix \ref{sec:Nemirovskii}.
On the other hand, part (B) has not been done before  to the extent that the
cut-off in the Biot-Savart equation
\eqref{Biot-Savart equation} and the Hamiltonian \eqref{ham Biot-Savart equation} would be rigorously justified and calculated, and this will be the main
goal of the derivations that follow.

The process of finding the Hamiltonian in terms of the vortex line is done in six steps. The result will be
\begin{equation}
\label{res1}
 H= \frac{\kappa^2}{16 \pi}\int_{|{\bf  s}-{\bf  s}'|>\xi_*} \frac{d{\bf  s} \cdot d {\bf  s}'}{|{\bf  s}-{\bf  s}'|} \,,
\end{equation}
where $\xi_* = 0.3416293 \pm 10^{-7}$. The value of the cut-off will be
shown to satisfy the following analytical formula:
\begin{equation}
\label{res2}
\xi_{*} = \frac{1}{2} \exp \left( -\frac{1}{2} - \int_0^\infty \left[\frac{d R(r)}{dr}\right]^2 \,r \,dr - \lim_{r\to\infty} \left[\int_0^r \frac{[R(r')]^2}{r'} \,dr' - \ln r\right] \right),
 \end{equation}
 where $R(r)$ is the vortex profile, solution of equation \eqref{eq:NLS-vortex}.\\

\noindent \textbf{Step 1: Obtaining the vortex-line Hamiltonian.} This was already done by writing
\eqref{eq:cons-Ef}.
We will assume no sound is present or excited by the moving vortices, which itself requires a justification if one wants to be fully rigorous. However, we expect this to be a difficult task akin to proving the validity of the balanced (wave-free) geophysical motions.
Physically, we could say that sound is not generated when the motions of the vortices are strongly subsonic, which is true when they are separated by the distances much greater than their core radii.

Thus, here we will simply postulate Hamiltonian   \eqref{eq:cons-Ef}
and assume that the density field rigidly follows the vortex cores, so that locally near the cores we have the Pitaevskii vortex profile. In this way, the last two terms in this Hamiltonian will be sub-dominant (after subtracting the constant contribution), but still  important because they affect the value of the cut-off length $\xi_*$.\\

\noindent \textbf{Step 2: Reformulating the problem in the form of a flow with constant density in the leading order.}
We do this by introducing a new ``velocity'' field
\begin{equation}\label{eq:v}
  {\bf  v} = \sqrt {\frac{\rho}{2}} \, {\bf  u}  \, 
\ ,
\end{equation}
so that the Hamiltonian is 
\begin{equation}
H =H_K + H_0\, , \qquad H_K = \frac{1}{2} \int 
   {v}^2
 \, d\mathbf{x}\,, \qquad  H_0 = \frac{1}{2} \int  [(\rho-1)^2  + 2|\nabla \sqrt{\rho}|^2 ]\, d\mathbf{x}\,,
\label{eq:cons-Ebsv}
\end{equation}
where $H_K$ formally looks like the kinetic energy of incompressible 
fluid with density equal to one. {The terms in $H_0$ depend on the vortex profile and contribute directly to the value of the cut-off length $\xi_*$. }

For an infinite straight (Pitaevskii) vortex, we have:
\begin{equation*}%\label{eq:v1}
  {\bf  v}  = \frac{\sqrt {2 \rho}}{r_\perp} \, \hat  {\! \boldsymbol \theta}
\ ,
\end{equation*} 
where $\, \hat {\! \boldsymbol \theta}$ is the unit vector in the azimuthal direction and $r_\perp$ is the distance to the vortex line.

Taking into account that that $\sqrt {\rho/2} \to \alpha\, r_\perp$, $\alpha=$~const,
when the distance to the vortex $r_\perp \to 0$, and that in the 
vortex $u = 2 /r_\perp$, we have
\begin{equation}
v \to 2 \alpha, \quad \hbox{as} \;\;\;  r_\perp \to 0.
\label{eq:v-as}
\end{equation}
Same relation holds for a more general vortex line when its curvature radius is much greater than the healing length.

We compute  $H_0$ explicitly in the case of the infinite straight vortex. This will be used later on. In terms of the vortex profile $R(r)$ we obtain, per unit length of vortex line, 
\begin{equation}
\label{eq:H_0_num}
\frac{H_0}{L} =  \frac{\kappa^2}{8\pi} \, \mu_0\,,
\end{equation}
where $\kappa$ is the quantum of circulation, $\kappa = 4 \pi,$ and
\begin{equation}
\label{eq:mu_0_def}
\mu_0 \equiv \int_0^\infty  \left[ \frac{1}{2} ([R(r)]^2-1)^2  + \left( 
\frac{dR(r)}{dr} \right)^2 \right]\,r\,dr .
\end{equation}
These integrals can be split as follows. First \cite{Roberts2003}, 
$$\int_0^\infty  ([R(r)]^2-1)^2\,r\,dr = 1.$$
Second, we obtain numerically
$$\int_0^\infty   \left( 
\frac{dR(r)}{dr} \right)^2 \,r\,dr  \approx 0.279090913.$$
Therefore 
$$\mu_0 \approx  0.779090913 \,.$$
\\

\noindent \textbf{Step 3: Rewriting the Hamiltonian 
\eqref{eq:cons-Ebsv} in such a way that the integrand functions are localised within the vortex cores.} This can be done at the expense of increasing dimensionality 
of the integral via performing an integration by parts followed
by applying the Helmholtz theorem:
\begin{equation}\label{3DBS-H}
H_K
= \frac 1 {8\pi} \int \frac{\boldsymbol
\omega ( {\bf  x}, t)
\cdot
\boldsymbol
\omega ( {\bf  x}', t) + \gamma( {\bf  x}, t)\, \gamma( {\bf  x}', t)
}{|{\bf  x} -  {\bf  x}'|}\  d{\bf  x} d{\bf  x}'   ,
\end{equation}
where 
$\boldsymbol
\omega ( {\bf  x}, t)$ is the ``$v$-vorticity'' field:
$\boldsymbol
\omega ( {\bf  x}, t) = \nabla \times {\bf v} ( {\bf  x}, t)$, and $\gamma( {\bf  x}, t)$ is the $v$-divergence field:
$\gamma( {\bf  x}, t) = \nabla \cdot {\bf v} ( {\bf  x}, t).$

To obtain this formula, we have had to use Helmholtz theorem on a domain that does not include the vortex lines, so that ${\bf v} ( {\bf  x}, t)$ satisfies the regularity conditions of the theorem. Now, it is possible to make this domain as close to the vortex lines as we like and show that the contribution of the remaining small vicinity of the vortex is vanishing because the singularity of ${\bf v} ( {\bf  x}, t)$ is mild enough.
So the integration domain in equation \eqref{3DBS-H} is the full $\mathbb{R}^3$. 

From here on, we will discard the contribution from the divergence terms $\gamma( {\bf  x}, t)$, because they are small if the vortex line's curvature radius is much greater than the healing length.

From \eqref{eq:v-as} we have asymptotics for the vorticity near the vortex centre:
\begin{equation}
\omega = \frac 1 r_\perp \partial_{r_\perp} (r_\perp v) \to
\frac { 2 \alpha} {r_\perp} , \quad \hbox{as} \;\;\;  r_\perp \to 0.
\label{eq:w-as}
\end{equation}
The vorticity field is strongly localised within the healing length from the vortex centre and rapidly decays at large distances from the centre.

We see that such a $v$-vorticity is no longer a delta-function distributed on the vortex centre, as it was the case for the original $u$-vorticity.
It is still singular at $r_\perp=0$, but the singularity is mild, and 
it leads to a non-divergent $H$---without a cut-off. The latter statement is true because the integral in $H$ is identical to the original convergent
integral \eqref{eq:cons-Ebs}.\\

\noindent \textbf{Step 4: Splitting the Hamiltonian in terms of local and far regions.} This is the key step. We introduce an intermediate length scale $a$ which is in between of the healing length $\xi$ and the curvature of
the vortex line $\ell$: $ \xi \ll a \ll \ell$. 
We split the integrals in $H_K$ as
$$
H_K = H_>+  H_<\ ,$$
where 
$$H_> =  \frac 1 {8\pi} \int_{|{\bf  x} -  {\bf  x}'| > a} \frac{\boldsymbol
\omega ( {\bf  x}, t)
\cdot
\boldsymbol
\omega ( {\bf  x}', t) }{|{\bf  x} -  {\bf  x}'|}\  d{\bf  x} d{\bf  x}'$$
and
$$H_< = \frac 1 {8\pi} \int_{|{\bf  x} -  {\bf  x}'| < a} \frac{\boldsymbol
\omega ( {\bf  x}, t)
\cdot
\boldsymbol
\omega ( {\bf  x}', t) }{|{\bf  x} -  {\bf  x}'|}\  d{\bf  x} d{\bf  x}',$$
so $H_>$ and $H_<$ are the integrals over the domains $|{\bf  x} -  {\bf  x}'| > a$
and $|{\bf  x} -  {\bf  x}'| \le a$ respectively. Notice that the term $H_0$ contributes to $H$ with a  term that is more in the spirit of $H_<$ than $H_>$ because the integrand in $H_0$ is supported in the vicinity of the vortex lines.

Writing  $H_>$ in the Biot-Savart form \eqref{ham Biot-Savart equation} comes cheaply because  one can replace $|{\bf  x} -  {\bf  x}'|$ in the denominator with $ |{\bf  s} -  {\bf  s}'|$, which would be valid up to $O(\xi/a)$ corrections. Then the transverse to the vortex line directions can be integrated out independently for the primed and the un-primed variables, and in the end we have:
\begin{equation}\label{eq:Ham>}
 H_> = \frac{\kappa^2}{16 \pi}\int_{|{\bf  s}-{\bf  s}'|>a} \frac{d{\bf  s} \cdot d {\bf  s}'}{|{\bf  s}-{\bf  s}'|} \ ,
\end{equation}
where $\kappa$ is the quantum of circulation, $\kappa = 4 \pi$.
The pre-factor here comes from the fact that the transverse integration of
$\boldsymbol
\omega ( {\bf  x})$ gives the circulation of ${\bf v} ( {\bf  x})$, i.e. $ 2^{3/2} \pi$.

To consider the $H_<$ integral, one can think of the vortex line as locally straight and use local  coordinates.
Let us consider two planes $A$ and $A'$ transverse to the vortex line and passing through the points ${\bf  x}$ and ${\bf  x}'$ respectively.
Let us introduce the change of the integration variables
 $({\bf  x}, {\bf  x}') \to ({ s_\parallel}, {\bf  r}_\perp , { s_\parallel}' ,{\bf  r}'_\perp) $, where ${ s_\parallel}$ and $ { s_\parallel}'$ are the lengths
along the vortex line to its intersections with  planes $A$ and $A'$;
 ${\bf  r}_\perp $ and ${\bf  r}'_\perp $ are the local 2D Cartesian coordinates within $A$ and $A'$
respectively. Taking into account that the contributions to the integral are limited to local regions,  $|{\bf  x} -  {\bf  x}'| \le a$, the change of variables amounts to a shift and rotation, i.e. its Jacobian is one, up to $O(a/\ell)$ corrections. The integration region is then  $|s_\parallel
 - s_\parallel'| \le a$,
and we have:
\begin{equation*}%\label{3DBS-H1}
H_<
=  \frac 1 {8\pi} \int_{|s_\parallel
 - s_\parallel'| \le a}  \frac{
\omega ( {r}_\perp)
 \omega ({r}'_\perp)
}{\sqrt{( s'_\parallel - s_\parallel)^2 +    
({\bf  r}'_\perp - {\bf  r}_\perp)^2}}\ d{ s_\parallel} d{\bf  r}_\perp d{ s'_\parallel} d{\bf  r}'_\perp   .
\end{equation*}
After integrating out ${\bf  r}_\perp$ and  ${\bf  r}'_\perp$, 
the remaining integrand will be a function of 
$s \equiv |s'_\parallel - s_\parallel|$ only (this $s$ should not be confused with $|{\bf s}|$). Also, up to $O(a/\ell)$ corrections
integral $H_0$ can be replaced with its value obtained for the straight line, equation \eqref{eq:H_0_num}. Combining $H_0$ and $H_<$, we get
\begin{equation}\label{3DBS-H2}
H_< +H_0
= \frac {\kappa^2 L} {8\pi} \mu_0 + \frac {\kappa^2} {16\pi} \int_{|s_\parallel
 - s_\parallel'| \le a}  f(| s'_\parallel - s_\parallel |) \ d{ s_\parallel}  d{ s'_\parallel}  = 
 \frac {\kappa^2 L} {8\pi } \left(\mu_0 + \int_0^{a}  f({s}) \ d{ {s}} \right)  ,
\end{equation}
where $\mu_0$ is a constant defined in equation \eqref{eq:mu_0_def}, $L$ is the total length of the vortex filament,
 and 
\begin{equation*}
%\label{eq:f-def}
f({s}) \equiv \frac{2}{\kappa^2} \int 
\frac{
\omega ( {r}_\perp)
 \omega ({r}'_\perp)
}{\sqrt{{s}^2 +    
({\bf  r}'_\perp - {\bf  r}_\perp)^2}} d{\bf  r}_\perp  d{\bf  r}'_\perp .  
\end{equation*}
To obtain the latter expression we performed a linear transformation from $(s_\parallel, s'_\parallel)$ to $(s'_\parallel - s_\parallel, s'_\parallel + s_\parallel)$, with Jacobian equal to $1/2$.\\

\noindent \textbf{Step 5: Defining the effective cut-off analytically and finding the effective Hamiltonian.}
At this point it is useful to define
the function
\begin{equation}
\label{eq:F-def}
F(a) \equiv \mu_0 + \int_0^{ a}  f({s}) \ d{ {s}}.
\end{equation}
At large distances, $ 1 \sim \xi \ll {s} \lesssim a$, we have
$ f({s}) \approx 1/{s}$
so that
\begin{equation}\label{eq:xi-def}
F(a) = \mu_0 + \int_0^{ a}  f({s}) \ d{ {s}} \approx \ln a + \mu_0 + C \equiv \ln (a/\xi_*) .
\end{equation}
This can be considered a definition of the effective cut-off length $\xi_*$ in the
Biot-Savart formulation, which is expected to be different from the
standard expression $\xi = 1/\sqrt{\rho_0} =1$ by an order-of-one factor. The task of finding $\xi_*$ numerically is left to the next step. Notice that at ${s}  \to 0$,  function $ f({s}) = o(1/{s})$, so the integral in 
\eqref{eq:xi-def} converges and $F(0)=\mu_0$. In fact, $f(0)$ is finite ($\approx 1.122$ by numerical estimation).

\begin{figure}[h]
\includegraphics[width=9cm]{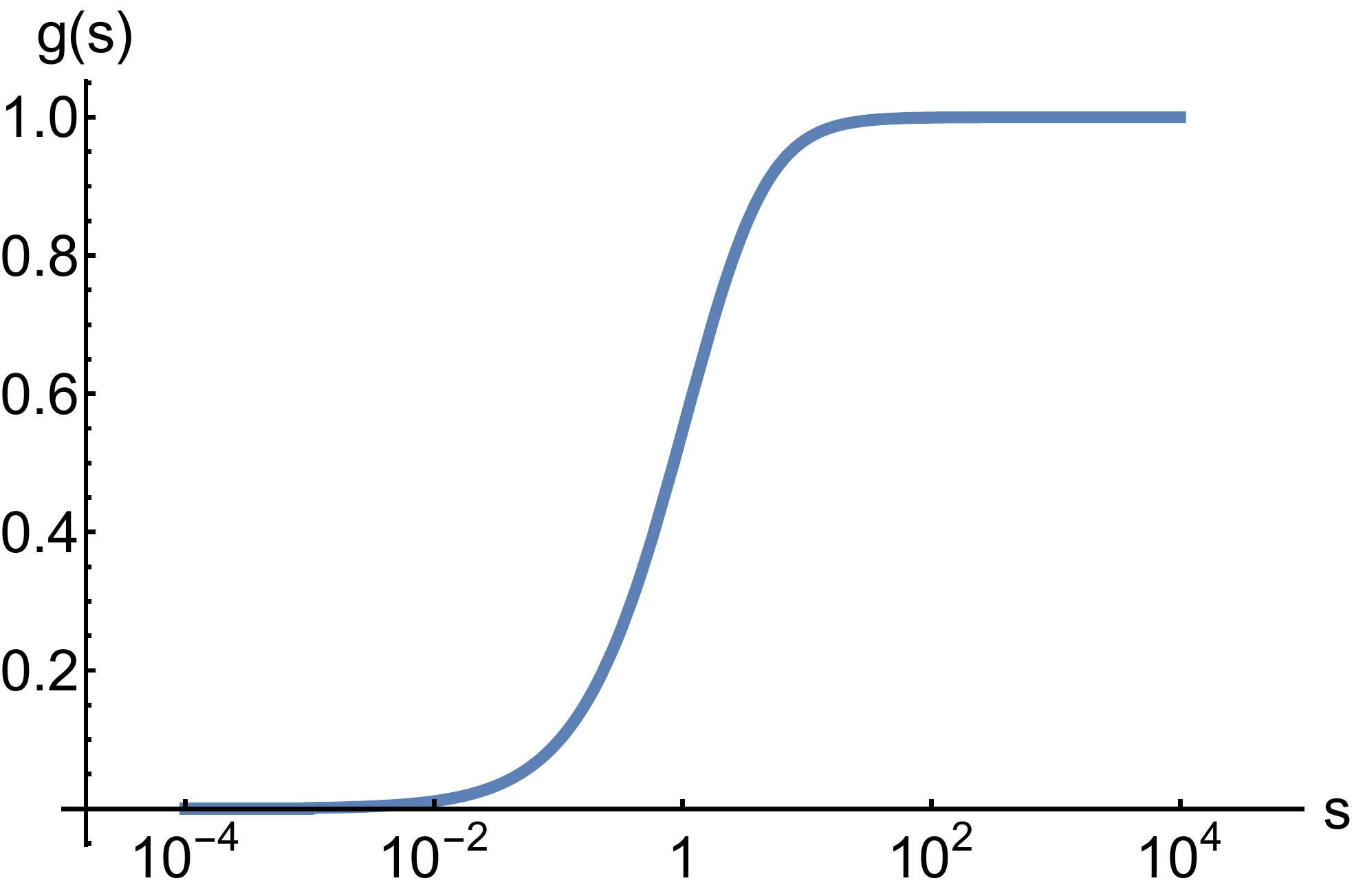}
\caption{
\label{fig:2} (Color online) Plot of numerical solution of function $g(s)$ defined in equation \eqref{eq:g-int1}.
}
\end{figure}

To obtain a closed formula for the full Hamiltonian, define the function 
\begin{equation}\label{eq:g-int}
g({s}) = {s}  f({s}) =   \frac{2 {s}}{\kappa^2} \int 
\frac{
\omega ( {r}_\perp)
 \omega ({r}'_\perp)
}{\sqrt{{s}^2 +    
({\bf  r}'_\perp - {\bf  r}_\perp)^2}} d{\bf  r}_\perp  d{\bf  r}'_\perp   
 ,
\end{equation}
which represents a ``smooth cut-off"; 
$
g(0) = 0, \;
g(\infty) = 1
$ (see figure \ref{fig:2}).

From equations (\ref{3DBS-H2}), (\ref{eq:F-def}) we rewrite
$$H_< +H_0
= \frac {\kappa^2 L} {8\pi} F(a)
= \frac{\kappa^2 L}{8 \pi}\int_{\xi_*}^{a} \frac{ds}{s} = \frac{\kappa^2}{16 \pi}\int_{a>|{\bf  s}-{\bf  s}'|>\xi_*} \frac{d{\bf  s} \cdot d {\bf  s}'}{|{\bf  s}-{\bf  s}'|}.$$

Adding this expression to   $H_>$ given in Eq.\eqref{eq:Ham>} we have:
\begin{equation*}%\label{eq:Ham}
 H= \frac {\kappa^2 L} {8\pi} \mu_0 + \frac{\kappa^2}{16 \pi}\int \frac{g(|{\bf  s}-{\bf  s}'|) \, d{\bf  s} \cdot d {\bf  s}'}{|{\bf  s}-{\bf  s}'|} 
= \frac{\kappa^2}{16 \pi}\int_{|{\bf  s}-{\bf  s}'|>\xi_*} \frac{d{\bf  s} \cdot d {\bf  s}'}{|{\bf  s}-{\bf  s}'|} \ ,
\end{equation*}
which is Eq.\eqref{res1} we were aiming to derive.
This is an asymptotically exact result valid when the vortex line's curvature radius is much greater than the healing length. \\

\noindent \textbf{Step 6: Finding the cut-off length $\xi_*$ numerically.}
To compute $g({s})$, we rewrite \eqref{eq:g-int} in polar coordinates:
\begin{equation*}%\label{eq:g-int1.0}
g({s}) =    \frac{{s}}{4 \pi} \int 
\frac{{r}_\perp {r'}_\perp 
\omega ( {r}_\perp)
 \omega ({r}'_\perp)
}{\sqrt{{s}^2 +    
{  r'}_\perp^2 +  {  r}_\perp^2 - 2 {r}_\perp {r}'_\perp \cos \theta}} \, d \theta d{  r}_\perp  d{  r}'_\perp.
\end{equation*}
The angular integration can be performed analytically, leading to an  expression for $g(s)$ that is easier to handle numerically. Dropping the `perp' symbols we get:  
$$g(s) =    \frac{{s}}{\pi} \int_0^{\infty}\int_0^{\infty} 
{r} \,  
\omega ( {r} ) \,{r'} 
 \omega ({r}' ) \frac{ K\left(\frac{4\, {r}  {r'} }{({r}  +{r'} )^2 + s^2}\right)
}{\sqrt{({r}  +{r'} )^2 + s^2}} \, d{  r}   d{  r}' ,
$$
where $K(m)$ is the complete elliptic integral of the first kind.
% ($m$ is called `parameter' and is equal to the square of the so-called `modulus' of the elliptic function).
Eqs.~\eqref{eq:v} and \eqref{eq:w-as} lead to  relation $r \,\omega(r) = \sqrt{2} \,\frac{dR(r)}{dr},$ where $R(r)$ is the solution of the ODE \eqref{eq:NLS-vortex} (in terms of the vortex density we have $R(r) = \sqrt{\rho(r)}$). Thus, we get:
\begin{equation}\label{eq:g-int1}
g(s) =    \frac{2\,{s}}{\pi} \int_0^{\infty}\int_0^{\infty} 
\frac{dR(r)}{dr}\,\frac{dR(r')}{dr'} \frac{ K\left(\frac{4\, {r}  {r'} }{({r}  +{r'} )^2 + s^2}\right)
}{\sqrt{({r}  +{r'} )^2 + s^2}} \, d{  r}   d{  r}' \,.
\end{equation}

We will compute this integral numerically using a very accurate numerical solution of equation \eqref{eq:NLS-vortex} for $R(r)$, combined with an asymptotic solution valid for large $r$, so that the error incurred in solving equation \eqref{eq:NLS-vortex} is kept uniformly below $5 \times 10^{-10}$ (see Appendix \ref{sec:accu_R}). 
Replacing the corresponding expression for $\frac{dR(r)}{dr}$ into \eqref{eq:g-int1}, the resulting 2D integral is computed numerically for several values of ${s}.$ The partial result is plotted in figure \ref{fig:2}.

\begin{figure}[h]
\includegraphics[width=9cm]{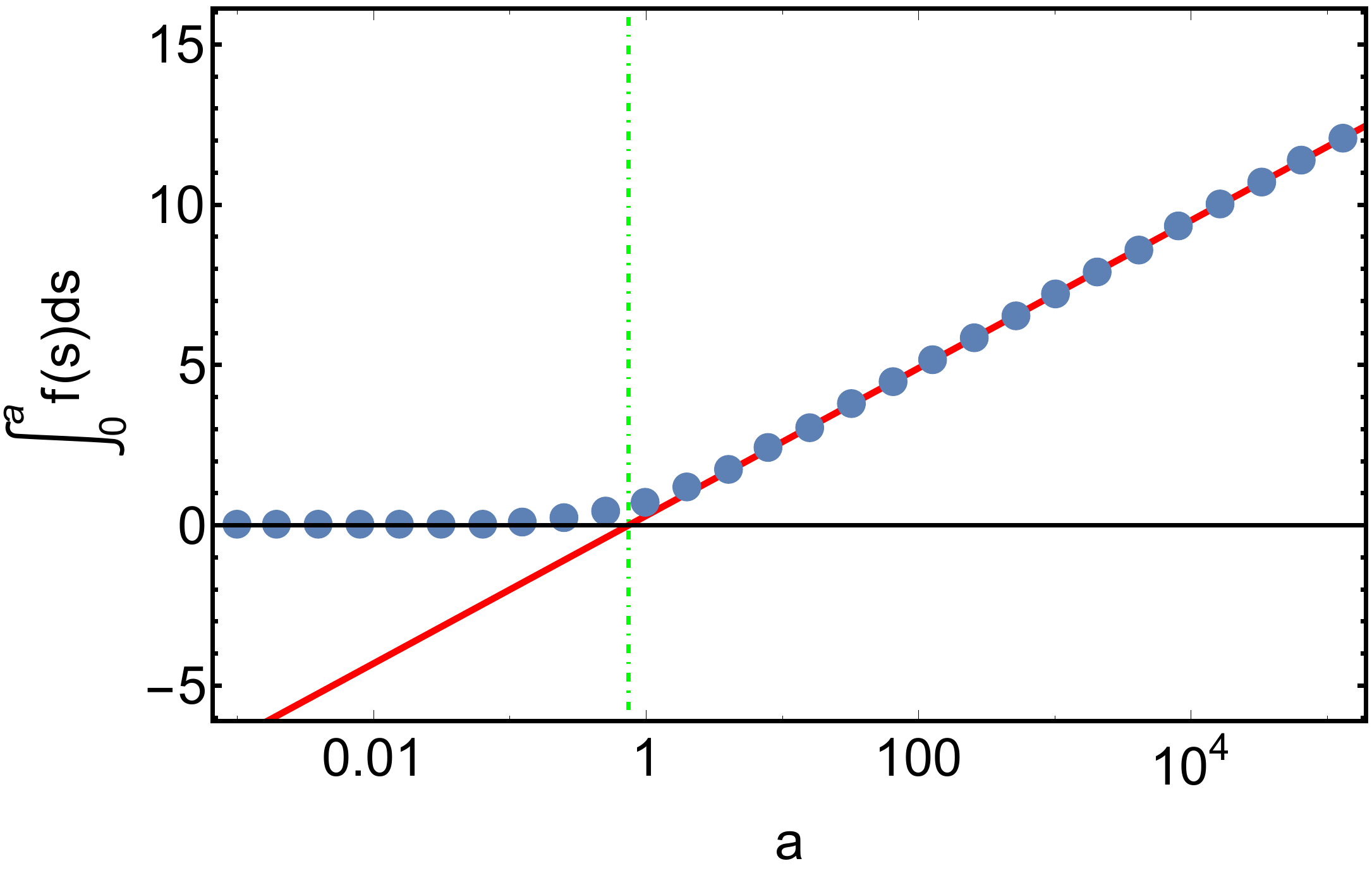}
\caption{\label{fig:3} (Color online) Blue filled circles: Numerical computation (in segments) of the integral $\int_0^a f(s)ds.$ The red oblique straight line corresponds to the fit $\int_0^a f(s)ds \approx \ln(a) + C.$ The green vertical dot-dashed line corresponds to $a = \exp(-C).$}
\end{figure}

To obtain the effective cut-off length $\xi_*$ we need to integrate numerically $f({s})$ $ (= g({s})/{s})$ over the range ${s} \in (0,a)$, thus obtaining $F(a)$ $ (= \mu_0 + \int_0^a f(s)ds),$ and compare the result with Eq.\eqref{eq:xi-def}. The numerical computation of $\mu_0$ is done accurately using the profile $R(r)$ and the result is $\mu_0 \approx 0.779090913.$  Figure \ref{fig:3} shows the results of numerically integrating $f({s})$ in segments, using \emph{Mathematica}'s global adaptive method with accuracy and precision goals of $10^{-10}$ each. The blue dots represent the values of $\int_0^a f(s)ds$ obtained numerically. The red straight oblique line corresponds to the fit 
$\int_0^a f(s)ds \approx \log(a) + C $ which is good in the asymptotic regime $a \gg 1$ (we went up to $a > 10^9$ to verify the asymptote). By looking at figure \ref{fig:3}, the value $\exp(-C)$ is by definition the value of $a$ at which the horizontal line and the red straight oblique line intersect. Combined with the numerical value of $\mu_0,$ the fit gives the following accurate estimation for the cut-off length:
\begin{equation}
\label{eq:xi*}
\xi_*  = \exp (- \mu_0 - C) = 0.3416293 \pm 10^{-7}.
\end{equation}
This result can be immediately generalised to the case $\rho_0 \ne 1$:
\begin{equation}
\label{eq:xi*'}
\xi_*   \approx 0.3416293 /\sqrt{\rho_0}.
\end{equation}

Finally, in Appendix \ref{sec:anal_formula} we derive the analytical formula \eqref{res2} that allows us to bypass the numerical fitting procedure. Using the accurate numerical solution for $R(r)$ on formula \eqref{res2} yields the same value for $\xi_*$ as in equation \eqref{eq:xi*'} above.

\section{Conclusions}

In this paper, for the first time, we have rigorously derived the Biot-Savart equation \eqref{res1} with cut-off \eqref{res2} from the 3D defocusing NLS equation \eqref{eq:NLS}. For this we assumed that the vortex line curvature and  the inter-vortex distance are much greater 
than the healing length. This setup corresponds to a very subsonic 
motion of the vortex lines, such that the density far from the vortices 
tends to a constant value ${\rho_0}$. {Correspondingly, we assumed that the core is rigid and has a fixed angle-independent profile. This is justified from the fact that all angle-dependent modes quickly dissipate via sound radiation \cite{Roberts2003}. Non-circular corrections to the core shape arising from curvature (e.g. for a vortex ring) are small and can be ignored in our derivations.}

We have found an 
accurate numerical value for the cut-off length, $\xi_*   \approx 0.3416293 /\sqrt{\rho_0}$. This value agrees with the common sense suggestion that it must be of the order of the healing length 
$\xi   = 1/\sqrt{\rho_0}$. However, the numerical coefficient shows an approximately threefold difference between these two quantities,
which may introduce significant corrections to previous estimates
of, e.g., the critical velocity for the vortex formation in a superfluid flow past an obstacle, critical distance for the vortex line reconnection process, speed of vortex rings, etc.

It is important to understand that the NLS model provides us with a significant advantage with respect to the ideal fluid model described by the Euler equation: the vortex profile is fixed in the former but not in the latter. As a result, the Biot-Savart equation would be much harder to derive from the Euler equation, and it would be unrealistic to expect that the result would be in a simple  form as in Eq.\eqref{res1} with a fixed
cut-off length. However, it would be interesting to explore possible justifications  of the Biot-Savart model extension with a variable cut-off length
physically corresponding to variability of the vortex radius due to the vortex stretching process. 

In the future, it would be interesting to conduct a comparative numerical study of the Biot-Savart and the NLS models for various physical
examples, including propagating and colliding vortex rings, Kelvin waves on vortex lines and reconnection of vortex lines. In particular, it would be interesting to see at what stage and under which conditions
the  Biot-Savart approach fails due to, e.g., close approach of different vortex lines to each other, appearance of regions with high curvature or loss of energy to acoustic waves.

\section{Acknowledgments}

MDB acknowledges support from Science Foundation Ireland (SFI) under Grant Number 12/IP/1491.

\appendix
\section{Accurate solution of vortex profile $R(r)$}

\label{sec:accu_R}

We implemented a very accurate shooting method to obtain $R(r)$ numerically  between $r=r_0 = \exp(-50) \approx 10^{-21}$ and $r=20$. The method uses \emph{Mathematica}'s stiffness switching solver with working precision of $22$ digits and accuracy and precision goals of $10^{-22}$ each. As a result, the absolute point-wise errors in solving equation \eqref{eq:NLS-vortex} is kept below $10^{-13}$ for all $r$ in the range $10^{-21} < r < 10^{-4},$ and below $10^{-17}$ in the range $10^{-4} < r < 20.$ The optimal value of derivative at $r_0$ is found to be 
$$\left[\frac{d R}{dr} \right]_{r=r_0}= 0.5831894958603292791791737551.$$
It turns out that this numerical solution cannot be continued much beyond $r=20$ without significant loss of accuracy. In order to continue the solution to higher values of $r$, we devised an asymptotic method. This method consists of the transformation 
$$R(r) = \exp(-Z(1/r^2)), $$
which leads to the ODE
$$4 q^2 Z''(q) - 4 q^2 Z'(q)^2 + 4 q Z'(q) - \frac{1-e^{-2 Z(q)}}{q} + 1=0.$$
This ODE is solved near $q=0$ in power series
$$Z(q) = \sum_{j=1}^{N} c_j q^j\,,$$
leading to an asymptotic solution that does not converge, but can be truncated so that the error in solving equation \eqref{eq:NLS-vortex} is kept as small as desired for $r$ large enough. In practice we found that $N=13$ is a good compromise. The coefficients $c_1, c_2, \ldots, c_{13}$ are, explicitly:
$$\frac{1}{2},\frac{5}{4},\frac{32}{3},\frac{1589}{8},\frac{64981}{10},\frac{989939}{3},\frac{168211250}{7},\frac{38006710085}{16},\frac{5510235057787}{18},$$
$$\frac{199454257136329}{4},\frac{110192683498843556}{11},\frac{14600012068277445755}{6}, \frac{9139380150115822460510}{13}.$$
 
Next, we patch this asymptotic series with the previous numerical solution by finding the intersection of the two functions. They intersect at 
$$r_{\mathrm{trans}} = 15.5752238612175596563309,$$
which defines a piecewise continuous solution at the transition point $r_{\mathrm{trans}}$. The jump in the derivative of this piecewise function is found to be reasonably small. In absolute terms,
$$\left|\left[\frac{d R}{dr} \right]_{r=r_\mathrm{trans}^+} - 
\left[\frac{d R}{dr} \right]_{r=r_\mathrm{trans}^-}
 \right| < 2\times 10^{-15},$$ 
while in relative terms,
$$\left|\left[\frac{d R}{dr} \right]_{r=r_\mathrm{trans}^+} - 
\left[\frac{d R}{dr} \right]_{r=r_\mathrm{trans}^-}
 \right| < 5 \times 10^{-12} \left[\frac{d R}{dr} \right]_{r=r_\mathrm{trans}^+}.$$ 
These errors are mostly due to the asymptotic solution's error in solving equation \eqref{eq:NLS-vortex}, which is of order $2 \times 10^{-10}$ at the transition point $r=r_{\mathrm{trans}}$ and rapidly decreases below $10^{-13}$ for $r>21.$
 
Details of the implementation of this accurate piecewise solution are found in the Supplemental Material \cite{suppl_mat}.

\section{Hamiltonian equations in terms of vortex lines}

\label{sec:Nemirovskii}

Below, we  reproduce derivations of Nemirovskii  \cite{Nemirovskii2004} 
simplifying them as appropriate for the NLS model \eqref{eq:NLS}.
Consistently with what we assumed when deriving $H$, we will
consider highly subsonic motions of the vortex lines which occur when
their curvature radius and mutual separations remain much greater 
than the healing length $\xi$. In this case one can neglect the acoustic 
waves and assume that function $\psi({\bf x},t)$ is fully determined by
the vortex line configuration ${\bf s}(\zeta, t)$:
$$
\psi({\bf x},t) \equiv \psi({\bf x}|{\bf s}(\zeta, t)).$$
In particular, the
time dependence in $\psi({\bf x},t)$ appears only implicitly via
${\bf s}(\zeta, t)$ so that
\begin{equation}
\label{eq:psi_t}
\partial_t \psi({\bf x},t) = \int \frac{ \delta\psi({\bf x}|{\bf s}(\zeta, t))}{\delta {\bf s}(\zeta', t)} \cdot  {\bf s}_t (\zeta', t) \, d \zeta'.
\end{equation}

Let us multiply equation \eqref{eq:NLS-h} by ${ \delta\psi({\bf x}|{\bf s}(\zeta, t))}/{\delta {\bf s}(\zeta_0, t)}$ add its complex conjugate and
integrate over the 3D physical space:
\begin{equation}
\label{A1}
i \int 
\partial_t \psi({\bf x},t) \frac{ \delta\psi^*}{\delta {\bf s}(\zeta_0, t)} \, d {\bf x} +c.c.
=  \int \left( \frac{ \delta H}{\delta \psi^*} \frac{ \delta\psi^*}{\delta {\bf s}(\zeta_0, t)} + \frac{ \delta H}{\delta \psi} \frac{ \delta\psi}{\delta {\bf s}(\zeta_0, t)}  \right)   d {\bf x} \equiv \frac{ \delta H}{\delta {\bf s}(\zeta_0, t)} .
\end{equation}
Dominant contribution to the integral on the LHS of this equation comes from
a small vicinity of the vortex line, where locally $\psi$ behaves as the Pitaevskii vortex \eqref{pv}, i.e.
\begin{equation*}
\psi({\bf x}|{\bf s}(\zeta, t)) \approx \psi_v({\bf x}_\perp) \equiv
\psi_v ({\bf s}(\zeta_\perp, t)- {\bf x}),
\end{equation*}
where $\zeta_\perp$ corresponds to the point on the vortex line 
which is closest to ${\bf x}$. Then we can write
\begin{equation*}
 \frac{ \delta\psi({\bf x}|{\bf s}(\zeta, t))}{\delta {\bf s}(\zeta', t)} =
\nabla_\perp \psi_v({\bf x}'_\perp) \, \delta(\zeta' - \zeta'_\perp).
\end{equation*}
Using this in the LHS of \eqref{A1} and its conjugate version in \eqref{eq:psi_t}, and integrating over 
$ d x_\parallel = |{\bf s}_{\zeta} (\zeta_0, t) | d\zeta_\perp $ and over
$\zeta'$
%and $\zeta_\perp$ 
using 
the delta-functions, the LHS of \eqref{A1} becomes:
\begin{equation*}
i   |{\bf s}_{\zeta} (\zeta_0, t) | \int (  {\bf s}_t (\zeta_0, t) \cdot \nabla_\perp \psi_v({\bf x}_\perp))
\nabla_\perp \psi_v^*({\bf x}_\perp) \, d {\bf x}_\perp
+c.c.
\end{equation*}
%where $ \dot {\bf s}(\zeta_0, t) \equiv \partial_t   {\bf s}(\zeta_0, t)$.
Using the vector triple product formula, we have:
$$
i \,(  {\bf s}_t (\zeta_0, t) \cdot \nabla_\perp \psi_v({\bf x}_\perp))
 \nabla_\perp \psi_v^*({\bf x}_\perp) +c.c.=
 i \,{\bf s}_t(\zeta_0, t) \times ( \nabla_\perp\psi_v^* ({\bf x}_\perp)
\times
\nabla_\perp \psi_v({\bf x}_\perp) )
$$
Substituting
 \eqref{pv}, we have
$$
 \nabla_\perp\psi_v^* ({\bf x}_\perp)
\times
\nabla_\perp \psi_v({\bf x}_\perp) 
=2i \frac{R(x_\perp) R'(x_\perp)}{x_\perp} \,
\hat {\bf t} ,
$$
where the prime here means derivative, ${ x}_\perp = |{\bf x}_\perp |$ and $\hat {\bf t} = \frac{{\bf s}_{\zeta} (\zeta_0, t) }{|{\bf s}_{\zeta} (\zeta_0, t) |}$ is a
unit vector tangential to the vortex line.
We have taken into account that 
$\nabla_\perp \theta = \hat {\boldsymbol \theta}/x_\perp$, where $\hat {\boldsymbol \theta}$ is a
unit vector in the azimuthal direction.

Thus, the LHS of \eqref{A1} becomes
\begin{equation*}
-  |{\bf s}_{\zeta} (\zeta_0, t) |  ( {\bf s}_t (\zeta_0, t) \times \hat {\bf t})
 \int_0^\infty 2\frac{R(x_\perp) R'(x_\perp)}{x_\perp} \, 2\pi x_\perp dx_\perp =
- 2\pi ( {\bf s}_t (\zeta_0, t)  \times {\bf s}_{\zeta} (\zeta_0, t)),
\end{equation*}
where we used the boundary conditions $R(0) = 0, R(\infty) = 1.$ Using this in \eqref{A1}, we finally obtain
\eqref{Ham BS equation}.

\section{Analytical expression of the cut-off length $\xi_*$ in terms of the 
vortex profile $R(r)$}

\label{sec:anal_formula}

Based on the  definition of $\xi_*$, we have:
$$\ln(1/\xi_*)  = \mu_0  + \lim_{a\to\infty} \left[\int_0^a f(s) ds - \ln(a) \right].$$
Using equation \eqref{eq:g-int1} and definition $f(s) = g(s) /s$ we get:
\begin{equation}
\label{eq:xi_temp}
\ln(1/\xi_*)  = \mu_0 + \lim_{a\to\infty} \int_0^\infty \int_0^\infty \frac{dR(r)}{dr} \frac{dR(r')}{dr'} \left[\frac{2}{\pi} \int_0^a \frac{ K\left(\frac{4\, {r}  {r'} }{({r}  +{r'} )^2 + s^2}\right)
}{\sqrt{({r}  +{r'} )^2 + s^2}} ds - \ln(a) \right] \, d{  r}   d{  r}' ,
\end{equation}
where we have used $\int_0^\infty \frac{dR(r)}{dr} dr = 1.$
Notice that it is known how to compute integrals of the form 
$$\int_0^\infty \frac{ K\left(\frac{4\, {r}  {r'} }{({r}  +{r'} )^2 + s^2}\right)
}{\sqrt{({r}  +{r'} )^2 + s^2}}\, s^{-\nu} \,ds,$$
after appropriate transformations that lead to tabulated integrals such as the ones found in \cite{Cv99}. The integral is finite for $1/2>\nu>0$ and has a simple pole at $\nu=0.$ Therefore we can regularise the integral over $s$ in equation \eqref{eq:xi_temp} by introducing a factor $s^{-\nu}.$ To keep the integral dimensionless, we choose the following dimensionless regularising factor: $\left(\frac{s}{r+r'}\right)^{-\nu}.$ One can then interchange the limits $a\to \infty$ and $\nu \to 0,$ provided we can cancel the pole at $\nu=0$ using a regularisation of the term $ \ln(a).$ The latter can be regularised as follows:
$$\ln(a) = \lim_{\nu \to 0^+} \left(\ln\left(\frac{r+r'}{2}\right)+\int_0^a \left(\frac{s}{r+r'}\right)^{-p\,\nu}\frac{1}{\sqrt{({r}  +{r'} )^2 + s^2}}ds \right),$$
where $p$ is a positive number, to be chosen in order to achieve the above-mentioned cancellation of the pole at $\nu=0.$
To achieve a more familiar presentation of the integral involving the complete elliptic function $K,$ we perform a standard transformation to obtain:
\begin{eqnarray*}
\ln(1/\xi_*)  = \mu_0 + \lim_{a\to\infty} \lim_{\nu\to 0^+} \int_0^\infty \int_0^\infty \frac{dR(r)}{dr} \frac{dR(r')}{dr'} &&\left[\frac{2}{\pi} \int_0^{\left(1+\frac{(r+r')^2}{a^2}\right)^{-\frac{1}{2}}} K\left(\frac{4\,r\,r'}{(r+r')^2} (1-k^2)\right) \frac{k^{-2\nu}dk}{(1-k^2)^{1-\nu}} \right.\\
 && \!\!\!\!\! \!\!\!\!\! \!\!\!\!\! \!\!\!\!\!  \left. - \ln\left(\frac{r+r'}{2}\right) - \int_0^a \left(\frac{s}{r+r'}\right)^{-p\,\nu}\frac{1}{\sqrt{({r}  +{r'} )^2 + s^2}}ds \right] \,dr\,dr'.
 \end{eqnarray*}

We now interchange the limits, obtaining first 
\begin{eqnarray}
\nonumber
\ln(1/\xi_*)  = \mu_0 +  \lim_{\nu\to 0^+} \int_0^\infty \int_0^\infty \frac{dR(r)}{dr} \frac{dR(r')}{dr'} &&\left[\frac{2}{\pi} \int_0^1 K\left(\frac{4\,r\,r'}{(r+r')^2} (1-k^2)\right) \frac{k^{-2\nu}dk}{(1-k^2)^{1-\nu}} - \ln\left(\frac{r+r'}{2}\right)\right.\\
\label{eq:xi_temp1}
 &&\left. - \int_0^\infty \left(\frac{s}{r+r'}\right)^{-p\,\nu}\frac{1}{\sqrt{({r}  +{r'} )^2 + s^2}}ds  \right] \,dr\,dr'.
 \end{eqnarray}

The integral involving the elliptic function is now in standard form, tabulated in \cite{Cv99}, Eqs.(4) and (1.a.4), with the result 
$$\frac{2}{\pi}\int_0^1 K\left(\frac{4\,r\,r'}{(r+r')^2} (1-k^2)\right) \frac{k^{-2\nu}dk}{(1-k^2)^{1-\nu}} = \frac{\sqrt{\pi}}{2 \cos \nu \pi} \frac{\Gamma(\nu)}{\Gamma(\nu+1/2)} \,{}_{2}F_{1}\left(\frac{1}{2},\nu;1;\frac{4\,r\,r'}{(r+r')^2}\right),$$
where ${}_{2}F_{1}$ is the hypergeometric function. As for the remaining integral, it amounts to
$$\int_0^\infty \left(\frac{s}{r+r'}\right)^{-p\,\nu}\frac{1}{\sqrt{({r}  +{r'} )^2 + s^2}}ds = \frac{\Gamma(p \nu/2) \,\Gamma(1/2-p\nu/2)}{2 \sqrt{\pi}}.$$
We see how the two integrals present simple poles at $\nu=0$ coming from the Gamma functions. Going back to Eq.\eqref{eq:xi_temp1}, we perform a Laurent expansion on the terms in the square brackets, obtaining
\begin{eqnarray}
\nonumber
\ln(1/\xi_*)  = \mu_0 + \lim_{\nu\to 0^+} \int_0^\infty \int_0^\infty \frac{dR(r)}{dr} \frac{dR(r')}{dr'} &&\left[ \frac{p-2}{2\,p\,\nu}  + \frac{1}{2} \frac{\partial}{\partial z}\left\{\,{}_{2}F_{1}\left(\frac{1}{2},z;1;\frac{4\,r\,r'}{(r+r')^2}\right)\right\}_{z=0} \right.\\
\nonumber
&&\left.- \ln\left(\frac{r+r'}{2}\right)+ {\mathcal O}(\nu) \right] \,dr\,dr'.
 \end{eqnarray}
We see that in order to cancel the pole we need to choose $p=2.$ Notice that the ${\mathcal O}(1)$-term that survives the limit does not depend on $p$, which indicates that the method is consistent. Therefore we obtain 
$$
\ln(1/\xi_*)  = \mu_0  +  \int_0^\infty \int_0^\infty \frac{dR(r)}{dr} \frac{dR(r')}{dr'} \left[\frac{1}{2} \frac{\partial}{\partial z}\left\{\,{}_{2}F_{1}\left(\frac{1}{2},z;1;\frac{4\,r\,r'}{(r+r')^2}\right)\right\}_{z=0} - \ln\left(\frac{r+r'}{2}\right)\right] \,dr\,dr'.
$$
A remarkable relation allows us to simplify this: the term in square brackets above is equal to
$
-\ln (\max(r,r')) + \ln 2.$
Also, replacing $\mu_0$ by its definition, equation \eqref{eq:mu_0_def}, we get
\begin{eqnarray*}
\ln(1/\xi_{*}) &=&  \frac{1}{2}+ \ln 2 + \int_0^\infty \left[\frac{d R(r)}{dr}\right]^2 \,r \,dr - \int_0^\infty \int_0^\infty \frac{dR(r)}{dr} \frac{dR(r')}{dr'} \ln (\max(r,r')) \,dr\,dr'.
\end{eqnarray*}
The latter double integral can be transformed to a single integral using integration by parts, 
\begin{equation*}
\ln(1/\xi_{*}) = \frac{1}{2}+\ln 2 + \int_0^\infty \left[\frac{d R(r)}{dr}\right]^2 \,r \,dr + \lim_{r\to\infty} \left[\int_0^r \frac{[R(r')]^2}{r'} \,dr' - \ln r\right],
 \end{equation*}
which is the analytical formula \eqref{res2} we have aimed to prove.
A numerical application of this formula using \emph{Mathematica} gives the same result as the previous fitting method, Eq.\eqref{eq:xi*}, namely   
$\xi_* = 0.3416293 \pm 10^{-7},$ which validates both the numerical methods and the analytical formula.

\bibliography{BS_NLS_references}

\end{document}